\documentstyle[11pt,aussois2003,twoside,epsf]{article}
\def\edcomment#1{\iffalse\marginpar{\raggedright\sl#1\/}\else\relax\fi}
\reversemarginpar
\newcommand{\gsim}{\raise 
        -2.truept\hbox{\rlap{\hbox{$\sim$}}\raise5.truept 
        \hbox{$>$}\ }}
\newcommand{\lsim}{\raise 
        -2.truept\hbox{\rlap{\hbox{$\sim$}}\raise5.truept 
        \hbox{$<$}\ }}  
\begin{document}
%
\title{Arc statistics with realistic cluster models}
%
\author{Massimo Meneghetti}
\affil{Dipartimento di Astronomia, Universit\`a di Padova, Vicolo
  dell'Osservatorio 2, I--35122, Padova, Italy}
\author{Matthias Bartelmann}
\affil{Max-Planck-Institut f\"ur Astrophysics,
  Karl-Schwarzschild-Strasse 1, D--85748 Garching, Germany} 
\author{Lauro Moscardini}
\affil{Dipartimento di Astronomia, Universit\`a di Bologna, Via
  Ranzani 1, I--40127, Bologna, Italy} 
\author{Elena Rasia}
\affil{Dipartimento di Astronomia, Universit\`a di Padova, Vicolo
  dell'Osservatorio 2, I--35122, Padova, Italy}
\author{Giuseppe Tormen}
\affil{Dipartimento di Astronomia, Universit\`a di Padova, Vicolo
  dell'Osservatorio 2, I--35122, Padova, Italy}
\author{Elena Torri}
\affil{Dipartimento di Astronomia, Universit\`a di Padova, Vicolo
  dell'Osservatorio 2, I--35122, Padova, Italy}

\label{page:first}
\begin{abstract}
Arc statistics is known to be a powerful cosmological tool. Numerical
lensing simulations show that orders of magnitude differences in the
number of {\em giant} arcs on the whole sky are expected in different
cosmological models. In this paper, we discuss the analytic and
numerical methods in arc statistics and show that analytic models fail
to reproduce the efficiency for strong lensing of more realistic
numerical cluster models. Then, we discuss two recent extensions of the
lensing simulations, i.e. the effects of cD galaxies in the lensing
clusters and the impact of cluster mergers on arc statistics. We show
that cD galaxies can increase the lensing cross sections for long and
thin arcs by perhaps up to $\sim50\%$, while major mergers can change
the cluster efficiency for producing such arcs by up to one order of
magnitude.   
\end{abstract}
\section{Introduction}
\label{section:arcstatistics}

Many factors determine the abundance of strong lensing events,
i.e. gravitational arcs, in galaxy clusters. Very shortly, since the
light deflection depends on the distances between observer, lens and
source, gravitational lensing depends on the geometrical
properties of the Universe. Moreover, gravitational arcs are rare
events caused by a highly nonlinear effect in cluster cores, and are
thus not only sensitive to the number density of galaxy clusters, but
also to their internal structure.

Since all of these factors depend on cosmology, the statistical study
of strong gravitational lensing events in galaxy clusters is a
powerful cosmological tool. In particular, we expect that arc
statistics is very sensitive to the values of $\Omega_0$ and
$\Omega_{\Lambda}$. 
Indeed, the expected number of {\em giant\/} arcs, usually defined as
arcs with a length-to-width ratio exceeding ten and apparent
$B$-magnitude less than $22.5$ (Wu \& Hammer, 1993), changes by orders
of magnitude between low- and high-density universes according to the
numerical models described in Bartelmann et al. (1998).

In the next sections of this paper, we discuss the possible methods
for constraining the cosmological parameters using arc statistics and
we consider some extensions of the previous studies, modeling in as
much detail as possible some of the effects that may influence or
distort the conclusions drawn from the morphology and number of
gravitationally lensed arcs.

\section{Numerical vs. Analytic methods}

Two approaches have been followed in arc statistics studies so far. The
first is based on numerical methods: the ray-tracing technique is used
for studying the lensing properties of clusters taken from N-body
simulations. This allows the most realistic description of the cluster
lenses because all effects which could play an important role in the
lensing phenomena (like asymmetries, substructures in the mass
distribution, etc.) are by construction taken into account. However,
given the long computation times required for full numerical
simulations of cluster lensing, it is currently not feasible to
perform such simulations for sufficiently many combinations of the
essential cosmological parameters, i.e.~the matter density parameter
$\Omega_0$ and the cosmological constant $\Omega_\Lambda$.

In a conceptually different approach, simple analytic, axially
symmetric models have been used for describing the density profiles of
cluster lenses. This method of investigation has the advantage that
the computation of the probability for arcs satisfying a specified
property is fast and can easily be performed for a continuous and wide
range of cosmological parameters, because the lensing properties of
these models are perfectly known and fully described by analytic
formulae. However they give a less realistic description of cluster
lenses.

The correspondence between analytic and numerical models remains
unclear. While the analytic studies by Cooray et al. (1999) and
Kaufmann \& Straumann (2000) find similar results as the numerical
simulations regarding the sensitivity of arc statistics to the cosmic
density parameter, their results are almost insensitive to the
cosmological constant, in marked contrast to the previously mentioned
results by Bartelmann et al. (1998), who found order-of-magnitude
changes in the expected number of giant arcs between low-density
models with and without a cosmological constant.

In order to check the consistency of the analytic and numerical
methods in arc statistics, we perform a comparative study of
lensing cross sections of numerical cluster models and several their
analytical approximations. We summarize here our main results
referring the reader to the paper by Meneghetti, Bartelmann \&
Moscardini (2003a) for a more detailed discussion.

\subsection{Axially symmetric models}

Previous analytic studies of arc statistics commonly used the Singular
Isothermal Sphere (SIS hereafter) model for describing cluster
lenses. The density profile of this model  is given by
\begin{equation}
  \rho(r)=\frac{\sigma_v}{2\pi G r^2} \ ,
  \label{equation:rhosis}
\end{equation}
where $\sigma_v$ is the velocity dispersion and $r$ is the distance
from the sphere center. 

This model is computationally convenient, but has an unrealistic
density profile for clusters. Indeed, several observations, for example
the occurrence of radial arcs in some galaxy clusters, indicate that
cluster density profiles are far from isothermal. 
Navarro, Frenk \& White (1997) (NFW hereafter) found that the
density profile of dark matter halos numerically simulated in the
framework of CDM cosmogony can be very well described by the radial
function
 \begin{equation}
   \rho(r)=\frac{\rho_{\rm s}}{(r/r_{\rm s})(1+r/r_{\rm s})^2} \ ,
   \label{equation:nfw}
 \end{equation}
within the wide mass range $3\times 10^{11} \lsim M_{\rm vir}/(h^{-1}
M_{\odot}) \lsim 3 \times 10^{15}$. The logarithmic slope of this
density profile changes from $-1$ at the center to $-3$ at large
radii. Therefore, it is flatter than that of the SIS in the inner part of
the halo, and steeper in the outer part. The two parameters $r_{\rm s}$
and $\rho_{\rm s}$ are the scale radius and the characteristic density of
the halo. They are not independent and can thus be expressed in terms
of the halo virial mass, which is in fact the only free parameter. 

An important feature of this model is that it reflects the
theoretically expected dependence of halo concentration, defined as
the ratio of the virial and scale radius, $c=r_{\rm vir}/r_{\rm s}$,
on cosmology. Halos are the more concentrated the earlier they form
and this important property is reproduced by the NFW model. 

Several different aspects of lensing by halos with NFW or generalized
NFW profiles can be found in Bartelmann (1996), Wright
\& Brainerd (2000), Li \& Ostriker (2002), Wyithe,
Turner \& Spergel (2001), Perrotta et
al. (2001), Meneghetti et al. (2003a),
Bartelmann et al. (2002a, 2002b). We refer the reader to those papers
for more details.

\subsection{Elliptical model}

The construction of lens models with elliptical or pseudo-elliptical
isodensity contours is generally quite complicated (Kassiola \&
Kovner, 1992; Kormann et al., 1994; Golse \& Kneib, 2002). Obtaining
the potential corresponding to these kinds of density distributions
can become extremely complicated, even for quite simple lens
models. It is simpler and often sufficient to model a lens by means of
an elliptical effective lensing potential.

For any axially symmetric lensing potential in the form
\begin{equation}
  \Psi(x)=f(x)
  \label{equation:lenpotgen}
\end{equation}
we can obtain its elliptical generalization by substituting
\begin{equation}
  x\rightarrow X=\sqrt{\frac{x_1^2}{(1-e)}+x_2^2(1-e)}\;,
\end{equation}
where $e=1-b/a$ is the ellipticity and $a$ and $b$ are the major and
minor axis of the ellipse. This ensures that the mass inside circles of
fixed radius remains constant as the ellipticity changes.

\subsection{Numerical models}

We analyze the lensing properties of five numerically simulated
cluster-sized dark-matter haloes, kindly made available by the GIF
collaboration (Kaufmann et al., 1999). They were obtained from $N$-body
simulations performed in the framework of three different cosmological
models. These are an Einstein-de Sitter model (hereafter SCDM); a
flat, low-density universe with a matter density parameter
$\Omega_0=0.3$ and a cosmological constant $\Omega_\Lambda=0.7$
($\Lambda$CDM); and an open, low-density model with $\Omega_0=0.3$ and
$\Omega_\Lambda=0$ (OCDM). The virial masses of the clusters at
redshift $z=0$ range between $\sim 5 \times 10^{14} M_{\odot}/h$ and
$\sim 2 \times 10^{15} M_{\odot}/h$. For each of these clusters, we
took several simulation snapshots at different epochs, between $z=0$ and
$z=1$.

The lensing properties of these clusters are studied using a
ray-shooting technique (Bartelmann et al., 1998, Meneghetti et al.,
2000, Meneghetti et al., 2001, Meneghetti et al., 2003a).  By
projecting the numerically simulated clusters along the line of sight,
we obtain surface density maps, which we use as lens planes in the
lensing simulations. A bundle of $2048 \times 2048$ light rays is then
traced through the mass distributions and their deflection angles are
computed by summing up the contribution from each mass element of the
deflector. We finally reconstruct the images of a large number of
background elliptical sources, placed at redshift $z_{\rm s}=1$.  It
results in a catalogue of simulated images which is subsequently
analyzed statistically.

\subsection{Lensing cross sections}

The lens efficiency for producing arcs with a given property is
quantified by the lensing cross section. By definition, it is the area
on the source plane where a source must be placed in order to be
imaged as an image with the specified property. 

For axially symmetric lenses, the lensing cross section can be
computed analytically or semi-analytically. For more complex lens
models, like the numerically simulated clusters or the
pseudo-elliptical NFW model, numerical techniques are required
(e.g. see Meneghetti et al., 2003a).

We focus here on long and thin arcs, i.e. arcs with a length-to-width
ratio $L/W$ larger than a minimal threshold. As anticipated earlier, we
compare the lensing cross sections for this kind of arcs of the numerical and
analytic models introduced in the previous sections. For clarity, we
present here the results obtained for the most massive halo only. The
behavior of the cross sections for the other numerical models is in
good qualitative and quantitative agreement with that obtained for
this cluster. 

The results are illustrated in Fig.~(1), where
the dotted line refers to the fully numerically simulated cluster,
while solid and dashed lines represent the cross sections of NFW and
SIS lenses, respectively, having the same virial mass as the numerical
cluster model. Finally, the shaded region in the same plot shows the
cross sections obtained by elliptically distorting the NFW lensing
potential with ellipticities $e$ in the range between $e=0.2$ and
$e=0.4$ (lower and upper limits, respectively). Results are shown for
$L/W\geq 7.5$ and they were obtained for the $\Lambda$CDM
model. Consistent results were found in the other investigated
cosmological models and for other minimal $L/W$ ratios.

\begin{figure}[t]
\plotfiddle{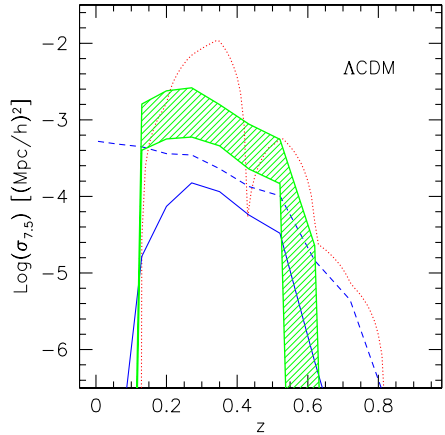}{5.5cm}{0.0}{30.0}{30.0}{-90.0}{-45.0}
\caption{Lensing cross sections for arcs with length-to-width ratio
  larger than $7.5$  as a function of the lens
  redshift. See text for more details.
}
\label{figure:comparison}
\end{figure}

The general trends in the lensing cross sections shown in Fig.~(1) can
be understood as follows. The strong-lensing efficiency of a mass
distribution depends on several factors. First, for the light coming
from the sources to be focused on the observer, the lens must be
located at a suitable distance from both the observer and the
sources. Second, the larger the (virial) mass of the lens is, the
stronger are the lensing effects it produces close to its
center. Finally, the more concentrated the lens is, the thinner are
the long arcs expected to be.  Since the lens mass grows with
decreasing redshift because new material is accreted by the halo and
deepens its potential well, the lensing cross sections are expected to
grow as well. On the other hand, when the lens is too close to the
observer, the cross section is geometrically suppressed, unless the
lens surface density profile is sufficiently steep and scale-free, as
for the SIS lenses. In fact, in this case the focusing by the lens is
strong enough to allow observers to see strongly distorted images of
background sources also in very near lenses, i.e.~at relatively small
redshifts.

The numerical models generally have much larger cross sections than
the analytic models. In particular, the cross sections for axially
symmetric NFW lenses are almost two orders of magnitude smaller. For
SIS lenses, the discrepancy with the numerical models is only
partially compensated by the unrealistically steep central density
profile, but the estimated values of $\sigma_{7.5}$ remain too
low. Introducing the elliptical distortion into the NFW lens model
allows the cross section to increase by roughly an order of magnitude
compared to the axially symmetric NFW model, but even then the
analytic cross sections fail to reproduce the numerical cross sections
unless unrealistically high values of $e$ are adopted.  


Since the ellipticity of the mass distribution alone cannot fully explain
the discrepancy between numerical and analytic models, the remaining
difference must be attributed to some factors which are missing from the
analytical models. The most important of those is certainly the
presence of substructure in the lensing mass distribution.

\begin{figure}[t]
\plotfiddle{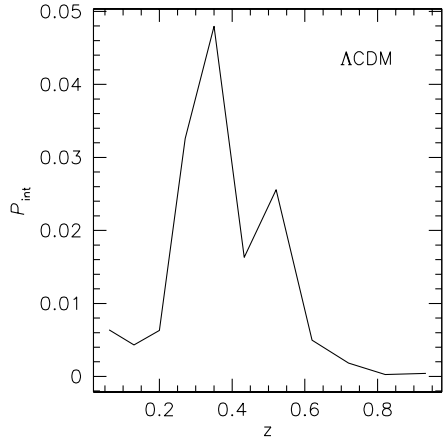}{5.5cm}{0.0}{30.0}{30.0}{-90.0}{-45.0}
\caption{Integrated multipole power as a function of redshift for the
  most massive cluster in our numerically simulated sample.}
\label{figure:comparison}
\end{figure}

Deviations of the projected mass distribution of a numerically
simulated cluster from the predictions of circular or elliptical
models can be quantified by means of a multipole expansion of its
surface density field (see Meneghetti et al., 2003a). By integrating
the power spectrum of this multipole expansion and by subtracting the
contributions from the monopole and the quadrupole, we build up a
quantity, $\bar{P}_{\rm int}$, which directly measures the amount of
substructures and the degree of asymmetry in the mass distribution of
the lens.

In Fig.~(2) we show how the integrated power $\bar{P}_{\rm int}$
averaged within the virial radius of the cluster changes as a function
of the cluster redshift for the lens whose cross section was plotted
in Fig.~(1). For better comparing the dependences on redshift of the
integrated power and the lensing cross sections, we have rescaled
$\bar{P}_{\rm int}$ with the effective lensing distance and with the
virial cluster mass. A quick comparison of this curve to the lensing
cross sections in Fig.~(1) shows that the redshifts where the
contributions of the dipole and the higher-order multipoles are
largest correspond quite well to those where the numerical cross
sections deviate most strongly from those of the elliptical models.

\section{The arc statistics problem}

In the previous section we showed that reliable predictions of arc
cross sections require realistic cluster models, i.e. numerically
simulated clusters. As anticipated earlier, by investigating the
lensing properties of a sample of clusters obtained from N-body
simulations in the framework of several cosmological models,
Bartelmann et al. (1998) found orders of magnitude differences between
the number of giant arcs which are expected to be seen on the whole
sky in different cosmological models. In particular, they estimated
that $\sim 3000$ giant arcs would be produced in a OCDM model, while
one order of magnitude less arcs are expected in a $\Lambda$CDM model.

Observations of the abundance of gravitational arcs in galaxy clusters
seem to be consistent only with the predictions for an open universe
(Luppino et al., 1999; Zaritsky et al., 2002, Gladders et
al., 2003). This is in pronounced disagreement with other observational
results, in particular those obtained from the recent experiments on
the cosmic microwave background (de Bernardis et al., 2000; Bennett et
al., 2003) and the observations of high-redshift type-Ia supernovae
(Perlmutter et al., 1998), which all suggest instead that the
cosmological model most favored by the data is spatially flat and
dominated by a cosmological constant. This is known as the {\em arc
statistics problem}.

Several extensions and improvements of the numerical simulations
failed in finding a solution to this problem in the lensing
simulations. For example, Meneghetti et al. (2000) studied the influence of
individual cluster galaxies on the ability of clusters to form large
gravitational arcs, finding that their effect is statistically
negligible.

In the next sections we discuss whether the presence of a central cD
galaxy or the occurrence of mergers in the lensing clusters can alter
the conclusions of previous arc statistics studies and solve the arc
statistics problem. 

\subsection{Effects of cD galaxies on arc statistics}

The centers of massive galaxy clusters are generally dominated by 
very massive ($\sim10^{13}\,M_\odot$) cD galaxies, which could in
principle noticeably affect the strong-lensing properties of their
host clusters. In fact, due to their more concentrated dark matter
halos, they may steepen the inner slope of the cluster density profile
and push the cluster critical curves to larger distances from the
center. Thus, the length of the critical curves may be increased, and
thus the probability for long arcs to form. Moreover, cD galaxies may
help their host clusters to reach the critical central surface density
for producing critical curves and becoming efficient strong lenses.

For investigating the effects of cD galaxies on arc statistics, we
study the lensing properties of the same sample of five numerically
simulated galaxy clusters described in Sect.~2.3, restricting our
analysis to the $\Lambda$CDM and the OCDM cosmological models
(Meneghetti, Bartelmann \& Moscardini, 2003b). We measure their
efficiency for producing tangential and radial arcs before and after
the inclusion of a cD galaxy. The central galaxy is modeled using
both the axially symmetric and elliptical models discussed in
Sect.~2.1 and 2.2 and assuming a range of virial masses and possible
orientations with respect to the mass distribution of the host
cluster.

The inclusion of the cD galaxy in the cluster lens has been done as
follows. Being linear function of mass, the total deflection angle of a ray
passing through a mass distribution is the sum of the contributions
from each mass element of the deflector. Therefore, in the case of a
galaxy cluster, we can decompose the cluster lens into its smoothed
dark matter component, plus the granular component contributed by its
galaxy population (see also Meneghetti et al.~2000). For both the
cluster and the galaxies, the main constituent is given by the dark
matter which forms their halos. Our model of the cluster containing a
cD galaxy can thus be fairly simple; we take the smoothed dark matter
distribution obtained from the numerical simulations described above,
and introduce a dark-matter halo resembling the galaxy. For each ray
traced through the lens plane, we compute the deflection angle by
summing the contributions from the cluster itself and the galaxy
haloes.

We first apply two axially symmetric models, namely spheres with the
NFW or the singular isothermal density profile. Second, we also apply
the pseudo-elliptical NFW lens model in order to account for the
possible elongation of the matter distribution of the cD galaxy. cD
galaxies typically appear to be of elliptical shape, with isophotal
axis ratios $b/a\sim0.8$ (Porter et al., 1991). Moreover, the
orientation of the brightest cluster ellipticals is usually not
random, but correlates well with that of their host
cluster. Asymmetries in the lensing matter distribution are known to
improve the ability of the cluster to produce long and thin arcs. We
also expect that the impact of a cD galaxy described by an elliptical
model is largest when its orientation is aligned with the elongation
direction of the host galaxy cluster. In the case of different
orientations, the cD galaxy tends to circularize the mass distribution
of the cluster in its central region. In order to quantify this
effect, we carry out two sets of simulations; in the first, we
randomly choose the orientation of the cD galaxy inside the cluster,
while in the second the orientation is chosen such that the directions
of the major and minor axes of the galaxy align with the major and
minor eigenvalues of the cluster's deflection angle field,
respectively. The galaxy ellipticity is assumed to be $e=0.2$.

In Fig.~3, we show the relative change of the lensing cross sections
for arcs with length-to-width ratio $L/W \geq 7.5$ as a function of
the cluster virial mass for the simulations in $\Lambda$CDM and OCDM
models. The lens redshift is $z_{\rm L}=0.27$, while sources are
placed at $z_{\rm S}=1$. In each panel, we plot the results for all
four models used to describe the cD galaxy. Both panels refer to
simulations in which the virial mass of the cD galaxy is
$5\times10^{13}\,h^{-1}\,M_\odot$.

As expected, the largest variations of the cross sections are
typically found if the cD galaxy is modeled as a pseudo-elliptical
NFW model whose orientation is aligned with that the host cluster. On
the other hand, cD galaxies with SIS profiles change the ability of
the numerical clusters for producing long and thin arcs only by a very
small amount.

\begin{figure}[t]
\plottwo{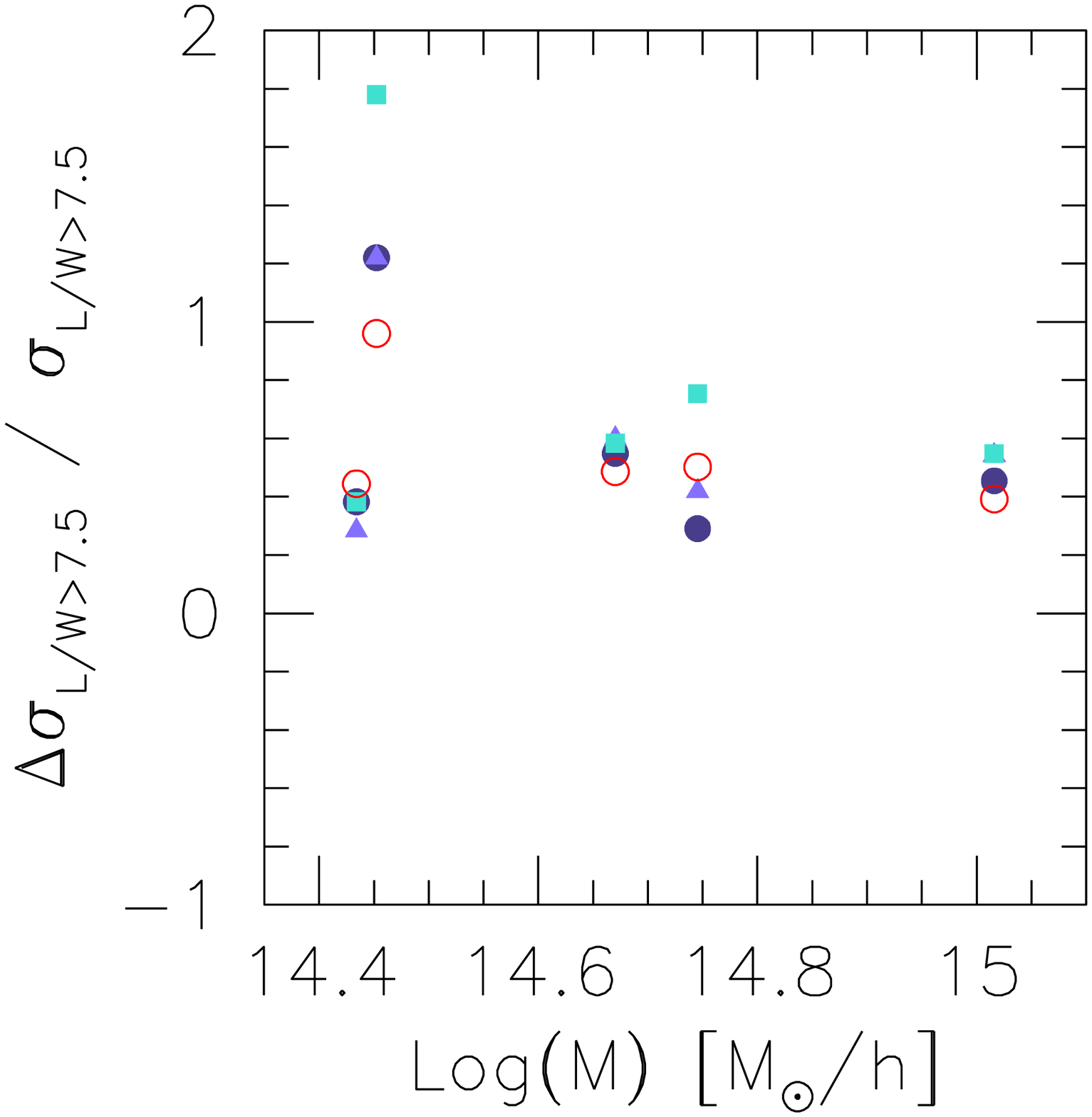}{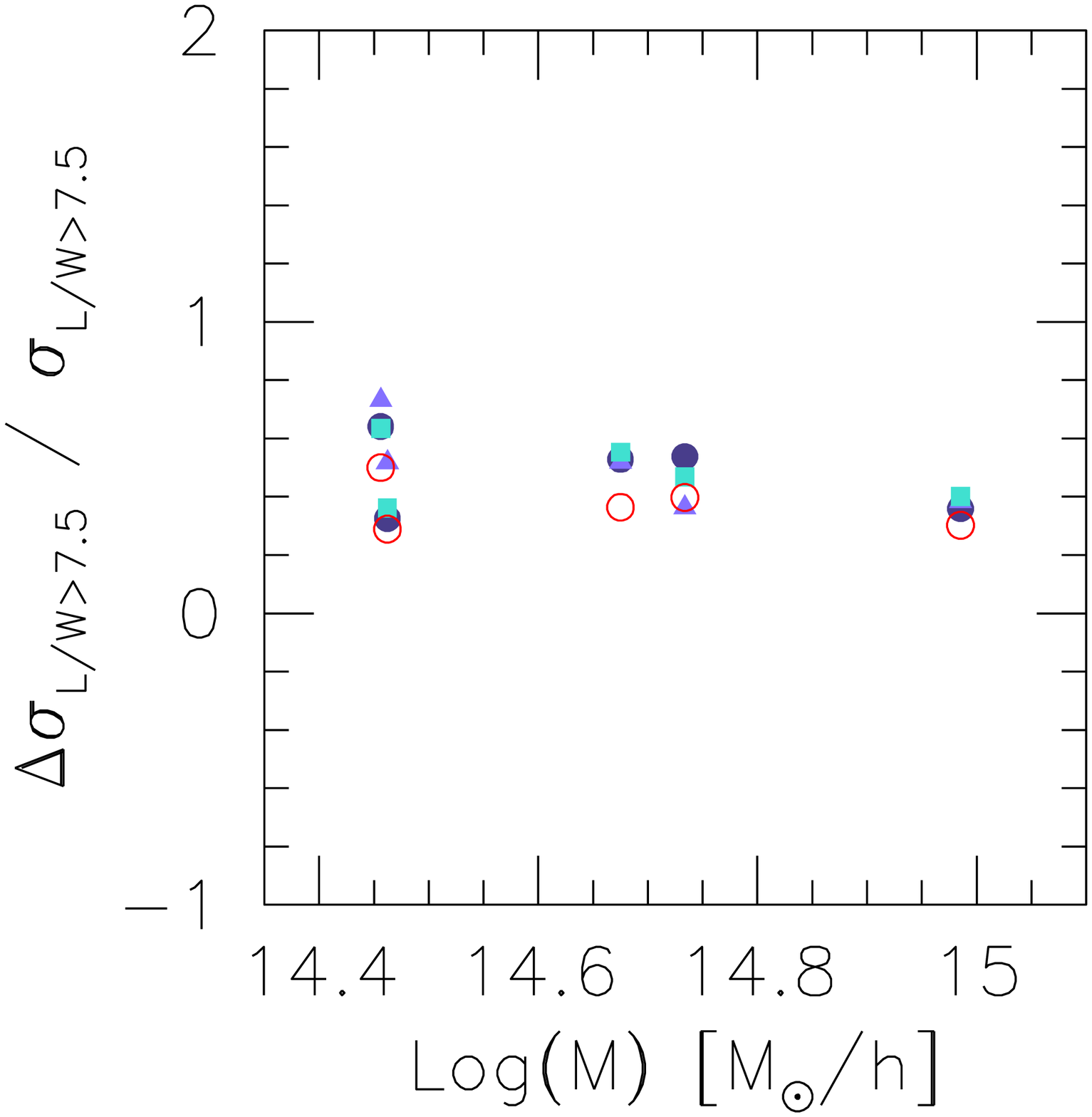}
\caption{Relative change in the cross sections for arcs with
  length-to-width ratios exceeding $7.5$ as a function of cluster mass
  for the numerically simulated clusters in the $\Lambda$CDM (left
  panel) and OCDM models. Filled circles, triangles and squares mark
  the results obtained modeling the cD as an NFW sphere, a
  pseudo-elliptical NFW model with random orientation, and aligned
  with the orientation of the host cluster, respectively; open circles
  show the results found modeling the cD galaxy as a SIS. The mass of
  the cD galaxy is $5\times10^{13}\,h^{-1}\,M_\odot$}
\label{figure:comparison}
\end{figure}    

In the $\Lambda$CDM model, a cD galaxy with mass $M_{\rm
cD}=5\times10^{13}\,h^{-1}\,M_\odot$ changes the lensing cross section
by a maximum amount between $60\%$ and $200\%$, depending on the total
cluster mass. The impact of the cD is generally larger in the less
massive clusters. Of course, less massive cD galaxies have smaller
impact on the lensing cross sections: in more realistic cases of cD
galaxies with mass $M_{\rm cD} \lsim 10^{13}\,h^{-1}\,M_\odot$,
tangential-arc cross sections are increased by not more than
$\sim50\%$.

A similar trend is found in the OCDM model, but the variations of the
cross sections are smaller in this case. For example, in the
simulations with the most massive cDs, the cross sections change by
approximately $40\%-80\%$ only. This behavior was expected because
the clusters in the OCDM model are generally more compact compared
than those in the $\Lambda$CDM model. Including the cD, the mass in
the very central part of the clusters changes less compared to the
clusters in the $\Lambda$CDM model. Cross sections for arcs with
different minimal length-to-width ratios show similar variations.

We thus conclude from our conservative estimates of the impact of cD
galaxies on strong-lensing cross sections by galaxy clusters that they
may increase the arc-formation probability by perhaps up to $\sim50\%$
in realistic situations, but certainly by far not enough for
explaining the discrepancy between simulations in $\Lambda$CDM models
and the observed abundance of arcs.

\subsection{The impact of mergers on arc statistics}

We investigate an other possible effect which could not
be properly considered in the previously mentioned numerical
simulations of gravitational lensing by galaxy clusters. In those
works, the lensing cross sections for giant arcs of each numerical
model were evaluated at different redshifts, with a typical time
separation between two consecutive simulation outputs of approximately
$\Delta t \sim 1$ Gyr. Therefore all the dynamical processes arising
in the lenses on time scales smaller than $\Delta t$ were not
resolved.

N-body simulations show that dark matter haloes of different masses
continuously fall onto rich clusters of galaxies (Tormen, 1997). The
typical time scale for such events is $\sim 1 \div 2$ Gyr, which
therefore might be too short for having been properly taken into
account in the previous lensing simulations.

Given the strong impact of substructures on the lensing properties of
galaxy clusters, it is reasonable to expect that during the passage of
a massive mass concentration through or near the cluster center, the
lensing efficiency might sensitively fluctuate. Indeed, when the
substructure is approaching the main cluster clump, the intensity of
the shear field and, consequently, the shape of the critical curves
might substantially change. Moreover, while the infalling dark matter
halo gets closer the cluster center, the projected surface density
increases, making the cluster much more efficient for strong lensing.

We use the ray-tracing technique for studying the
lensing properties of a numerically simulated galaxy clusters while a
mass concentration orbit very close to the cluster center. In
particular, we investigate an ``optimal'' projection of this
cluster, where a substructure is seen to pass exactly through the
center of main cluster clump. 

The numerical model we study here is part of a set of 17 objects
obtained using the technique of re-simulating at higher resolution a
patch of a pre-existing cosmological simulation. The re-simulation
method is described in Tormen et al. (1997). A detailed discussion of the
dynamical properties of the whole sample of these simulated clusters
is presented elsewhere (Tormen, Moscardini \& Yoshida 2003, in
preparation). A major merger occurs in this cluster between redshifts
$z=0.25$ and $z=0.15$. At redshift $z \sim 0.25$, when their viral
regions merge, the main cluster clump and the infalling substructure
have virial masses of $\sim 7 \times 10^{14} h^{-1} M_{\odot}$ and
$\sim 3 \times 10^{14} h^{-1} M_{\odot}$, respectively. In order to
have a very good time resolution to resolve in detail all the merging
phases, we decided to re-simulate the cluster between $z=0.25$ and
$z=0.15$, obtaining 101 equispaced outputs which we use for our
following lensing analysis.  

\begin{figure}[t]
\plotone{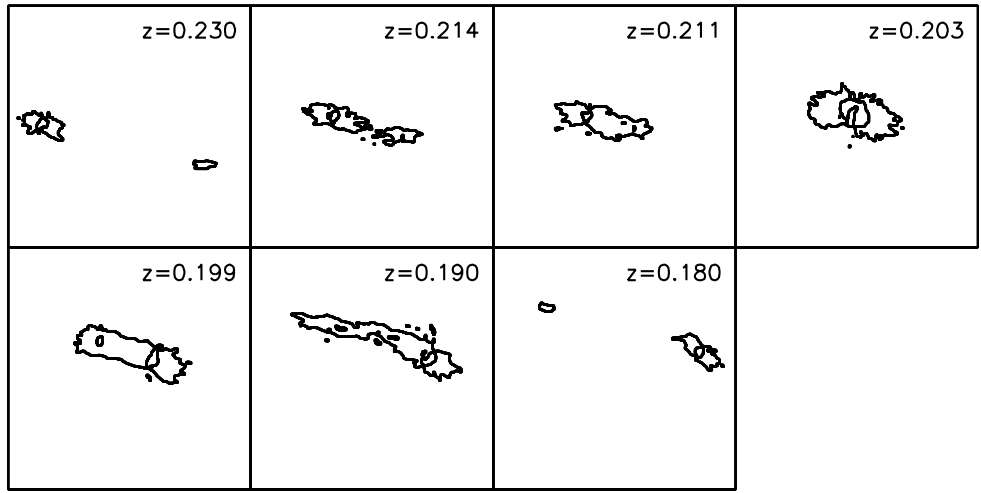}
\caption{Critical lines of the numerically simulated galaxy cluster at
  the several redshifts between $z=0.230$ and $z=0.180$, during the
  merging phase. The scale of each panel is $375''$.}
\end{figure}

As expected, the critical lines of the numerical lens evolve strongly
during the merger event. We show the critical lines at some relevant
redshifts in Fig.~(4). At redshift $z=0.230$, the main cluster clump
and the infalling substructure develop separate critical lines. The
largest mass concentration also produces a small radial critical line
(enclosed by the more extended tangential critical line). While the
merger proceeds, the tangential critical lines are stretched towards
each other. This is due to the increasing shear in the region between
the mass concentrations. The critical lines merge approximately at
redshift $z=0.214$. After that, there exists a single critical line,
which, after a short phase of shrinking, expands isotropically while
the two clumps overlap. The isotropic expansion is due to the
increasing convergence due to the larger amount of matter concentrated
at the cluster center. This happens at $z\sim 0.203$. When the
substructure moves to the opposite side, the tangential critical line
stretches again and reaches its maximum elongation at $z \sim
0.190$. Then, separate critical lines appear around each clump. Their
size decreases for $z \rightarrow 0$ because both the shear and the
convergence between the two mass concentration decrease as their
distance grows.

The lensing cross sections for long and thin arcs change during the
merger accordingly to the evolution of the critical lines. As an
example, the cross section for arcs with $L/W \geq 7.5$ as function of
redshift is shown in Fig.~(5). The cross section grows by a factor of
two between $z\sim 0.240$ and $z \sim 0.220$. Then, it further
increases by a factor of five between $z\sim 0.220$ and $z\sim 0.200$,
i.e. within $\sim 0.2$ Gyr. The curves have three peaks, located at
redshifts $z_1=0.214$, $z_2=0.203$ and $z_3=0.190$. The peaks at $z_1$
and $z_3$ correspond to the maximum extent of critical curves along
the merging direction {\em before} and {\em after} the moment when the
merging clumps overlap; the peak at $z_2$ occurs when the distance of
the infalling substructure from the merging clump is minimum. Two
local minima arise between the three maxima at redshifts $z_4=0.211$
and $z_5=0.199$, where the cross sections are a factor of two smaller
than at the peaks. At these redshifts, the critical lines have shrunk
along the merging direction. The cross section reduces by more than
one order of magnitude after $z=0.190$.
 
\begin{figure}[t]
\plotfiddle{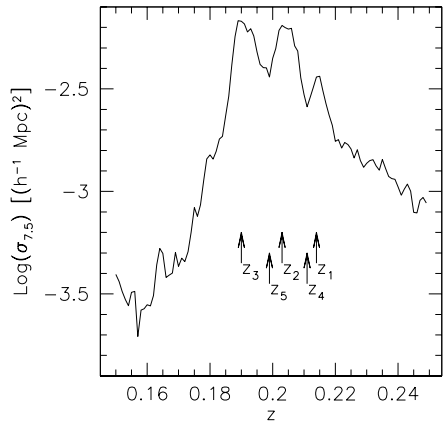}{5.5cm}{0.0}{30.0}{30.0}{-90.0}{-45.0}
\caption{Lensing cross section for arcs with $L/W>7.5$ as function of
  redshift.}
\end{figure}

Therefore, during the merger, our simulated cluster becomes extremely
more efficient to produce tangential arcs. The infalling substructure
starts affecting the cross sections for long and thin arcs when its
distance from the main cluster clump is approximately equivalent to
the cluster virial radius ($\sim 1.5 \ h^{-1}$Mpc), and the largest
effects are seen at three different times: 1) when the critical lines
(and the corresponding caustics) merge, i.e. when the shear between
the mass concentrations is sufficient to produce the largest
elongation of the critical lines along the direction of merging. This
happens $\sim 100$ Myr before and after the substructure crosses the
cluster center; 2) when the two clumps overlap, i.e. when the
projected surface density or convergence is maximal, producing the
largest {\em isotropic} expansion of critical lines and caustics.

Therefore, our results show that cluster mergers could play an
important role for arc statistics. In particular, since the lensing
efficiency grows by one order of magnitude during mergers, they might
be the solution of the {\em arc statistics problem}.

\begin{figure}[t]
\plotfiddle{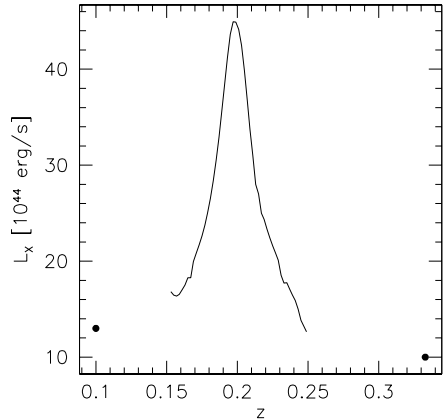}{5.5cm}{0.0}{30.0}{30.0}{-90.0}{-45.0}
\caption{X-ray luminosity $L_x$ of the numerically simulated cluster as
  function of redshift. The curve is accurately sampled in the redshift
  range $z=0.15 \div 0.25$. The X-ray luminosity has been measured
  also at $z=0.1$ and $z=0.333$, where is indicated by the filled circles.}
\end{figure}

It is quite important to notice that mergers might have some other
important observational implications to account for.
In fact the largest sample of clusters used for arc statistics studies
(Luppino et al, 1999) was selected in the X-ray band, where the luminosity
is due to bremsstrahlung emission. This is very sensitive to the
dynamical processes arising in the cluster, since it is proportional
to the square gas density. Therefore, we expect that the cluster X-ray
luminosity has large variations during a merging phase.

By using a code for simulating observations in the X-ray band by the
{\em Chandra} satellite (Gardini et al., 2003, in preparation), we
measure the observed X-ray luminosity of our numerical cluster at
several times during the merging phase.  In Fig.~(6) we show the
observed X-ray luminosity as function of redshift. The curve has a
narrow and almost symmetric peak located at $z\sim 0.200$. The X-ray
luminosity grows by more than a factor of four between $z\sim 0.300$
and $z\sim 0.200$, by roughly a factor of $\sim 2.5$ between $z\sim
0.230$ and $z\sim 0.200$ and by roughly a factor of $\sim 1.55$
between $z\sim 0.210$ and $z\sim 0.200$. The width at half maximum of
the peak is approximately $\Delta z \la 0.05$.

If a cluster sample is built by collecting all the objects with X-ray
luminosity $L_X$ larger than a given minimum threshold, we expect that
many merging clusters are present among them, since they are stronger
X-ray emitters.  Since these are all very efficient clusters for
producing gravitational arcs, this could introduce a bias in the
observationally determined frequency of {\em giant arc}, which could
become so large with respect to what predicted by previous numerical
lensing simulations in $\Lambda$CDM model.  However, it is quite
difficult to make more robust conclusions here since only one cluster
has been analyzed. Further investigations are needed on this subject.
In any case, our results show that much caution must be used when
selecting clusters for arc statistics studies through their X-ray
emission.

\section{Conclusions}
In the first part of this paper, we discussed the analytic and
numerical methods used in arc statistics studies. We showed that both
axially symmetric and elliptical analytic models fail to reproduce the
efficiency for producing long and thin arcs of numerically simulated
galaxy clusters. The deviations of the numerical lensing cross
sections from their analytical approximations can be attributed to
substructures inside clusters and tidal fields contributed by the
cluster surroundings, effects which cannot reasonably and reliably be
mimicked in analytic models. On the basis of these results, we
conclude that, in order to derive precise constraints on cosmology or
cluster structure and evolution, realistic cluster lens models are
required, for which numerical simulations seems to be the only
reliable choice. 

In the second part, we discussed two recent extensions of the
numerical lensing simulations. First, we investigate the effects of cD
galaxies on the cluster efficiency for long and thin arcs. We found
that reasonably massive cD galaxies at the cluster center may increase
the arc-formation probability by perhaps up to $\sim50\%$. Second, we
study the impact of major mergers on arc statistics. By measuring with
high time resolution the cross sections for long and thin arcs of a
numerical cluster model during a merger phase, we verified that the
cluster ability of producing tangential arcs may grow by one order of
magnitude while such events occur. This result is particularly
important because it confirms that mergers in clusters might play a
very important role in arc statistics. In particular they might be a
possible solution to the well know arc statistics problem.

\label{page:last}
\end{document}